\documentclass[aps,prl,showpacs,notitlepage,superscriptaddress,floatfix,twocolumn]{revtex4-1}
\usepackage{bbm}
\usepackage{mathrsfs}
\usepackage{epsfig}
\usepackage{graphicx}
\usepackage{amsfonts}
\usepackage[figuresright]{rotating}
\usepackage{amssymb}
\usepackage{amsmath}
\usepackage{dcolumn}
\usepackage{bm}
\usepackage{braket}
\usepackage{comment}
\usepackage{mathtools}
\usepackage{xcolor}
\usepackage{multirow}
\usepackage{setspace}

\def\avg#1{\langle#1\rangle}

\def\be{\begin{equation}} \def\ee{\end{equation}}
\def\bea{\begin{eqnarray}} \def\eea{\end{eqnarray}}

\def\nn{\nonumber}

\usepackage{color}

\usepackage[normalem]{ulem}

\begin{document}
	\title{Pairing symmetry and topological surface state in iron-chalcogenide superconductors}
	\author{Lun-Hui Hu}
	\affiliation{Department of Physics, University of California,
		San Diego, California 92093, USA}
	\author{P. D. Johnson}
	\affiliation{Condensed Matter Physics and Materials Science Division,
		Brookhaven National Laboratory, Upton, New York 11973}
	\author{Congjun Wu}
	\affiliation{Department of Physics, University of California,
		San Diego, California 92093, USA}
	\begin{abstract}
		The symmetries of superconducting gap functions remain an important question of iron-based superconductivity.
		Motivated by the recent angle-resolved photoemission spectroscopic measurements
		on iron-chalcogenide superconductors, we investigate the influence of
		pairing symmetries on the topological surface state.
		If the surface Dirac cone becomes gapped in the superconducting phase, it
		implies magnetization induced from time-reversal symmetry breaking pairing
		via spin-orbit coupling.
		Based on the crystalline symmetry constraints on the Ginzburg-Landau free
		energy, the gap function symmetries are among the possibilities of
		$A_{1g(u)}\pm iA_{2g(u)}$, $B_{1g(u)}\pm iB_{2g(u)}$, or, $E_{g(u)}\pm i E_{g(u)}$.
		This time-reversal symmetry breaking effect can exist in the
		normal state very close to $T_c$ with the relative phase between
		two gap functions locked at $\pm \frac{\pi}{2}$.
		The coupling between magnetization and superconducting gap functions
		is calculated based on a three-orbital model for the band structure of iron-chalcogenides.
		This study provides the connection between the gap function
		symmetries and topological properties of the surface state.
	\end{abstract}
	\maketitle
	
	The discovery of iron-based superconductors \cite{kamihara_jacs_2008} opened
	a new direction in the study on unconventional superconductivity \cite{sigrist1991,vanharlingen1995,tsuei2000}.
	Significant progress has subsequently been made in searching for
	new superconductors \cite{chen_nature_2008,chen_prl_2008,ren_epl_2008,rotter_prl_2008, sasmal_prl_2008,ni_prb_2010,nakai_prl_2010,wang_ssc_2008,
		borisenko_prl_2010,mizuguchi_apl_2008,hsu_pnas_2008,fang_prb_2008,
		chen_prb_2009,guo_prb_2010,qian_prl_2011},
	and their pairing mechanisms have attracted considerable
	attention \cite{chubukov_arcmp_2012,wang_science_2011,chen_nsr_2014,book_johnson2015,si_nrm_2016}.
	The parent compounds are metals with multiple Fermi surfaces around
	both $\Gamma$- and $M$- points.
	The possibility of the fully gapped extended $s_{\pm}$-wave superconducting
	gap function is supported by various experimental evidence and theoretical
	calculations \cite{mazin_prl_2008, kuroki_prl_2008, parker_prb_2008,
		chubukov_prb_2008,bang_prb_2009,seo_prl_2008,chen_prl_2009,
		chen_prl_2009,wang_prl_2009,hu_prb_2015}.
	On the other hand, several theoretical studies suggest that $s_{\pm}$ and 
	$d_{x^2-y^2}$ pairings are nearly degenerate in the iron-pnictide superconductors \cite{kuroki_prl_2008,graser_njp_2009}, leading to the possibility of a novel time-reversal (TR) symmetry breaking
	pairing $s_\pm+id_{x^2-y^2}$ \cite{lee_prl_2009}.
	It was proposed that a resonance mode carrying the $B_{1g}$-symmetry \cite{lee_prl_2009,scalapino2009}, which can be detected via Raman
	spectroscopy \cite{wen2014}, exists if the $s_\pm$ and $d_{x^2-y^2}$
	pairings are nearly degenerate.
	TR symmetry breaking pairing also naturally arises in mixed singlet
	and triplet pairing states \cite{wu2010,hinojosa_prb_2014}.
	A chiral $d+id$ pairing state is also found to spontaneously generate 
	the gaps of the Haldane model \cite{brydon2018}.

	Recently, the topological band structure of iron-based superconductors has
	aroused a great deal of attention.
	The FeSe$_{1-x}$Te$_x$ family with a wide range of composition $x$ is of
	particular interest \cite{hsu_pnas_2008,fang_prb_2008,
		yeh_epl_2008,sales_prb_2009,li_prb_2009,lcc_prb_2015,wang_prb_2015,wu_prb_2016,
		johnson_prl_2015,rameau2019}.
	Recent evidence shows that FeSe$_{0.45}$Te$_{0.55}$ is a strong topological insulator exhibiting a single Dirac cone on the (001) surface
	\cite{zhang_science_2018}.
	With lowering the temperature below $T_c=14.5K$, both bulk and the surface become superconducting \cite{miao_prb_2012,yin_nat_phy_2015}.
	The surface superconductivity is predicted to be topologically
	non-trivial \cite{wu_prb_2016,wang_prb_2015,xu_prl_2016}.
	Excitingly, in the same system, Majorana zero modes in vortex cores have been observed \cite{wang_science_2018,kong_nsr_2018} exhibiting
	the signature of spin-selective Andreev reflection \cite{sun_prl_2016,hu_prb_2016,he_prl_2014}.
	Similar evidence to vortex core Majorana modes is also
	observed in (Li$_{0.84}$Fe$_{0.16}$)OHFeSe system \cite{liu_arxiv_2018}.
	
	However, the non-trivial topology of the surface superconductivity in FeSe$_{0.5}$Te$_{0.5}$ is mostly a property inherited from the normal
	state band structure in a similar way to the Fu-Kane proposal of
	two-dimensional topological superconductivity via the proximity effect \cite{fu_prl_2008}.
	It does not directly reveal the symmetry properties of the superconducting
	gap functions.
	It would be highly desirable if the topological surface states could be used for phase-sensitive detections to unconventional pairing symmetries
	\cite{vanharlingen1995,tsuei2000}.
	In contrast, in a very recent laser-based angle-resolved photoemission
	spectroscopy (APERS) experiment on FeSe$_{0.3}$Te$_{0.7}$ \cite{johson_expt},
	the surface Dirac cone is observed to develop a gap as the system enters the superconducting state.
	The surface Dirac point is well-below the Fermi energy.  Thus the splitting
	cannot be the superconducting gap, 
	but implies TR symmetry breaking in the spin channel, directly
	correlated with the superconducting transition.
	In earlier literature, both TR symmetry breaking pairing of the types $s+id$ \cite{lee_prl_2009,platt_prb_2012,khodas_prl_2012,
		fernandes_prl_2013,kang_prb_2018} and $s+is$ \cite{stanev_prb_2010,marciani_prb_2013,maiti_prb_2013,ahn_prb_2014}
	have been proposed.
	However, in both cases magnetization only appears around impurities.
	
	In the present work, we investigate how the topological surface states are affected
	by TR breaking gap functions, which in turn constrains the possible pairing symmetries.
	Note that the degenerate Dirac cone in the surface state is protected by TR symmetry 
	in the normal state due to the non-trivial band structure topology. 
	Nevertheless, if the superconducting state breaks TR symmetry, the degeneracy of 
	the Dirac cone is no longer protected in the superconducting state.
	By employing the Ginzburg-Landau formalism, we explore possible TR
	breaking gap functions which can induce magnetization via spin-orbit coupling
	to split the degeneracy at the surface Dirac point.
	Based on crystalline symmetry analysis, the superconductivity
	gap symmetries include the possibilities of $A_{1g(u)}\pm iA_{2g(u)}$, $B_{1g(u)}\pm iB_{2g(u)}$, and $E_{g(u)}\pm i E_{g(u)}$.
	In the normal state sufficiently close to $T_c$, the relative phase
	between two gap functions can still be locked at $\pm\frac{\pi}{2}$
	even though neither of them is long-range ordered.
	Calculations based on a three-orbital model are performed to derive the
	coupling between magnetization and superconducting gap functions.
	In doing so, our study bridges the topological properties of
	the surface state and the pairing symmetries of the superconducting gap functions.
	
	\begin{figure}[!htbp]
		\centering
		\includegraphics[width=0.7\linewidth]{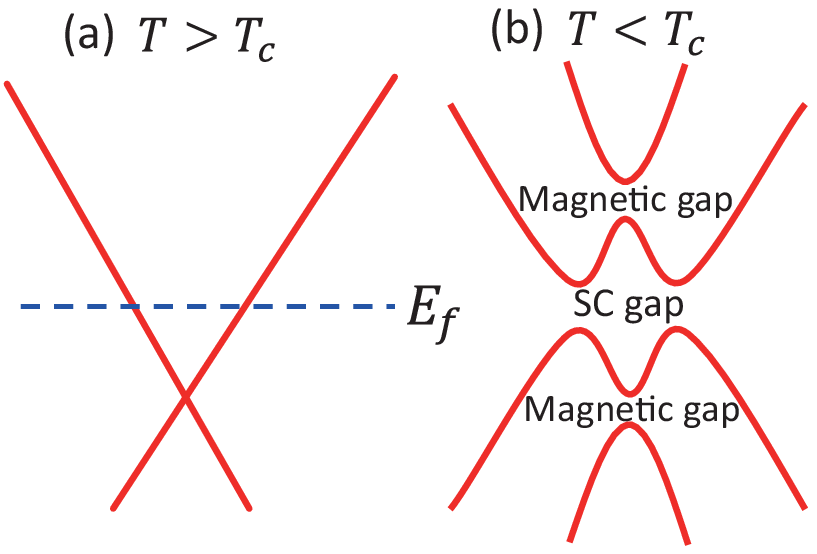}
		\caption{Sketch of the surface spectra near the $\Gamma$-point.
			($a$) In the  normal state ($T>T_c$), due to the non-trivial band structure
			topology, a surface Dirac cone develops inside the bulk band gap.
			($b$) The surface state spectra in the superconducting state ($T<T_c$).
			Two gaps appear:  the superconducting gap at the chemical potential $\mu$,
			and the splitting of the Dirac cone is due to the magnetic ordering
			breaking TR symmetry.
		}
		\label{fig-bands}
	\end{figure}
	
	We begin with a discussion of the splitting of the surface Dirac cone
	in the superconducting state.
	In the FeSe$_{1-x}$Te$_x$ materials, the degeneracy of the surface Dirac
	point is protected by the band topology if the normal state maintains
	TR symmetry.
	Consequently a surface Dirac cone appears at the $\Gamma$-point, as shown in
	Fig. \ref{fig-bands} ($a$), described by an effective $k\cdot p$
	Hamiltonian $H_{sf}=v(k_x\sigma_y - k_y\sigma_x)-\mu$, where
	$\sigma$'s are Pauli matrices defined for the Kramers doublet at
	the $\Gamma$-point, and  $\mu$ is the chemical potential.
	A superconducting gap $\Delta$ by itself, {\it i.e.},
	$H_\Gamma=-\mu +\Delta w_- +\Delta^* w_+$ where $w_\pm=w_x\pm i w_y$
	are Pauli matrices in the Nambu space of the particle-particle channel,
	does not lift the degeneracy.
	To split the degeneracy, a mass term breaking TR symmetry is necessary,
	{\it i.e.}, $\Delta H_\Gamma= -m_z \sigma_z$,  and the associated
	Bogoliubov spectra become $\pm m_z +\sqrt{|\Delta|^2+\mu^2}$
	around the $\Gamma$-point.
	The sketch of the Bogoliubov dispersion is illustrated in
	Fig.~\ref{fig-bands}($b$) with both particle and hole branches.
	The mass term corresponds to a splitting between two eigenstates of
	$\sigma_z$.
	Since no magnetic field is applied, $m_z$ should arise from the Weiss
	field of a ferromagnetic
	ordering along the $z$-axis induced by superconductivity.

	Now we examine how the ferromagnetic order $m_z$ can be induced in the
	superconducting state.
	Apparently, this requires the spontaneous breaking of TR symmetry.
	Indeed, within the Ginzburg-Landau (GL) formalism \cite{volovik1985,sigrist1991,sigrist2005,lee_prl_2009,wu2010},
	it has been shown that the mixing between two gap functions $\Delta_{1,2}$,
	which are TR invariant by themselves and possess different pairing
	symmetries, leads to the spontaneous TR symmetry breaking.
	The corresponding physical consequences were studied
	in the case of iron-based superconductors.
	$\Delta_{1,2}$ cannot form a symmetry invariant at the quadratic level,
	but they do at the quartic level via
	\bea
	F_4=\beta |\Delta_1|^2 |\Delta_2|^2 +\beta^\prime
	(\Delta_1^{*,2} \Delta_2^{2}+c.c).
	\label{eq:GL}
	\eea
	The $\beta^\prime$-term locks the relative phase between two gap functions, which
	equals $\beta^\prime |\Delta_1|^2|\Delta_2|^2 \cos2 \Delta \varphi$, where $\Delta \varphi=\varphi_1-\varphi_2$,
	and $\varphi_{1,2}$ are the phases of two gap functions.
	When $\beta^\prime>0$, $\Delta\varphi$ is pinned at $\pm\frac{\pi}{2}$,
	giving rise to complex gap functions $\Delta_1\pm i \Delta_2$, which break
	TR symmetry spontaneously.
	This formalism also applies to the case that $\Delta_{1,2}$ form a
	two-dimensional (2D) irreducible representation.
	The complex gap functions $\Delta_1\pm i \Delta_2$ distribute more evenly
	over the Fermi surface than the real ones $\Delta_1\pm \Delta_2$, hence,
	they are energetically more preferable at the mean-field level
	\cite{wu2010}.
	The corresponding Cooper pairs carry non-zero
	orbital moments, which could generate magnetic fields at boundaries
	as shown earlier \cite{volovik1985,sigrist2005}.
	However, these magnetic fields are typically of the order of 1 Gauss,
	for which the Zeeman energy is negligible.
	Instead, here we consider the spin magnetization $m_z$ coupling to
	$\Delta_{1,2}$ through a cubic term as
	\bea\label{eq-mz-sc1-sc2}
	\mathcal{F}_{M}= \alpha m_z^2 +
	i\gamma m_z\left(\Delta_1\Delta_2^\ast - \Delta_1^\ast\Delta_2 \right),
	\label{eq:GL-mag-sc}
	\eea
	This satisfies both the $U(1)$ and TR symmetry.
	$\alpha>0$ is assumed in Eq.~\eqref{eq:GL-mag-sc}, and hence there is no spontaneous
	magnetic ordering by itself, rather, the magnetization is induced,
	$m_z = \frac{\gamma}{\alpha} |\Delta_1\Delta_2| \sin \Delta \varphi$,
	{\it i.e}, by coupling to the TR breaking superconducting orders.
	The sign of $m_z$ is determined by the relative phase $\Delta\varphi$ between
	$\Delta_{1,2}$.

\begin{figure}[!htbp]
		\centering
		\includegraphics[width=0.7\linewidth]{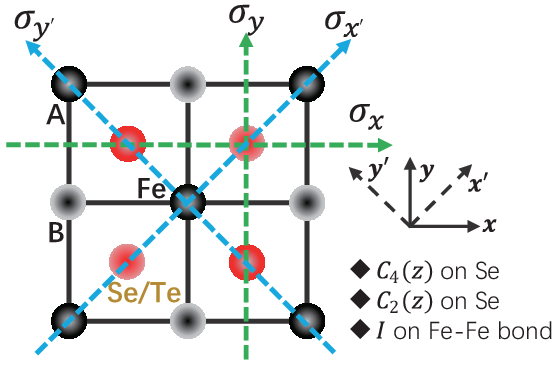}
		\caption{Sketch of the lattice structure of a trilayer FeSeTe unit.
			The $z$ axis is perpendicular to the plane.
			The $P4/nmm$ space group contains the $C_{4v}$ point group
			symmetries centering at Se(Te) anions, the $D_{2d}$ symmetries
			centering at Fe cations, and the inversion symmetries $I$
			with respect to the Fe-Fe bond centers.
		}\label{fig-lattice_symmetry}
	\end{figure}

	The free energy density ${\cal F}_M$ of Eq.~\eqref{eq:GL-mag-sc} is further
	required to satisfy all crystalline symmetries.
	At elevated temperatures the pristine FeSe and FeTe crystals are layered quasi-2D systems, whose
	Bravais lattices are primitive tetragonal.
	They exhibit a tri-layer structure with each unit cell consisting of two
	Fe cations and two Se(Te) anions:
	A square lattice of Fe cations in the middle layer sandwiched between two
	layers of Se(Te) anions in a $\sqrt 2 \times \sqrt 2$ structure.
	The Se(Te) lattices above and below the iron planes are off-set
	by one Fe-Fe bond length, and their projections are at iron plaquette centers.
	The crystalline space group is the non-symmorphic one $P4/nmm$ \cite{fang_prb_2008,li_prb_2009,miao_prb_2012,yin_nat_phy_2015},
	which is reviewed in the Supplemental Material (S.M.) I \cite{supp_mat}.
	It can be decomposed into 16 cosets: 8 of them are denoted as
	$g_i T $ where $T$ is the translation group
	of the primitive tetragonal lattice and
	$g_i (i=1\sim 8)$ span the point group $C_{4v}$
	centering at Se(Te) anions, and the other 8 cosets are
	$Ig_i T$ by further applying the inversion $I$ with respect to the
	Fe-Fe bond center.
	In the actual experimental systems of FeSe$_x$Te$_{1-x}$, the distribution of
	Se and Te breaks the $P4/nmm$ symmetry, nevertheless, this effect is
	weak after averaging over random configurations and will be neglected below.
	As shown in Fig. \ref{fig-lattice_symmetry}, the rotations with respect to
	Se/Te and Fe are 4-fold and 2-fold denoted as $C_{4}(z)$ and $C_{2}(z)$,
	respectively.
	The point group symmetry centering around the Fe cations is $D_{2d}$.
	The vertical reflection planes are along $x$, $y$ denoted as
	$\sigma_{x},\sigma_{y}$, and are along the diagonal lines $x^\prime$
	and $y^\prime$ denoted as $\sigma_{x^\prime}$ and $\sigma_{y^\prime}$, respectively.
	
	We consider the order parameter properties under the crystalline
	symmetries.
	In multi-orbital systems, the superconducting gap function is expressed as
	\bea
	\Delta=\sum_{\mathbf{k}}  \tau_{ij} f^l(\mathbf{k})  M^l_{ab}
	\Delta_{bj,ai}(\mathbf{k}),
	\label{eq:gapfunc}
	\eea
	where repeated indices mean summation;
	$a,b$ refer to the orbital band components, and $i,j$ are the sublattice
	indices of two Fe-cations in one unit cell;
	$\tau$ is a $2\times 2$ matrix representing the sublattice channel;
	$f^l(\mathbf{k})$ is the angular form factor of momentum $\mathbf{k}$,
	and $M^l$ is the pairing matrix in the orbital channel;
	$l$ is the index for multiple combinations between
	$f^l(\mathbf{k})$ and  $M^l$.
	The pairing matrix is defined as
	$\Delta_{bj,ai}(\mathbf{k})=\sum_{\mathbf{k}}
	i\sigma_{y,\alpha\beta} \avg{|c^\dagger_{\mathbf{k},\alpha b j}  c^\dagger_{-\mathbf{k},\beta a i}|}$,
	where $i\sigma_y$ projects out the singlet pairing
	with Greek indices representing spin components, and $\avg{||}$ represents
	averaging over the thermal equilibrium state.
	
	As required by Fermi statistics, for spin-singlet pairing, the product
	$\tau_{ij} f(\mathbf{k}) M_{ab} $ in Eq.~\eqref{eq:gapfunc} needs to
	be even under the combined operations of $\mathbf{k}\to -\mathbf{k}$
	and the transposes of $M$ and $\tau$.
	$f(\mathbf{k})$ can be an even function taking the forms of $1$,
	$\cos k_x \pm \cos k_y$,  $\cos k_x \cos k_y$, $\sin k_x\sin k_y$,
	$\cos (k_x\pm k_y)$.
	or, an odd function among $\sin k_x$, $\sin k_y$ and $\sin(k_x\pm k_y)$.
	We choose the three $t_{2g}$-orbital bases, $d_{x^\prime z}$, $d_{y^\prime z}$
	and $d_{xy}$, where $d_{x^\prime z} (d_{y^\prime z})$ extends along the diagonal
	$x^\prime (y^\prime)$ direction as depicted in S. M. II \cite{supp_mat}.
	Hence, $M$ is a $3\times 3$ Hermitian matrix which is expanded
	in terms of the Gell-mann matrices $\lambda_{i}
	(i=1\sim 8)$ under the basis in the sequence of
	$(d_{x^\prime z},d_{y^\prime z}, d_{xy})$, and the $3\times 3$
	identity matrix $\lambda_0$, whose expressions are presented in S. M. II \cite{supp_mat}.

	The representation of a gap function under the crystalline symmetry
	group is determined by the symmetry properties of $f(\mathbf{k})$,
	$M$, and $\tau$ as analyzed and presented in the S. M. II \cite{supp_mat}.
	Their possible symmetries are denoted as $A_{1g(u)}$, $A_{2g(u)}$,
	$B_{1g(u)}$, $B_{2g(u)}$, and $E_{g(u)}$, respectively, where
	$g$ and $u$ represent even and odd parities, respectively.
	The $A$, $B$, and $E$ symbols represent the discrete angular momentum,
	loosely speaking, they are analogues to the $s$, $d$, and $p$-wave
	symmetries, respectively.
	$A_1$ and $A_2$ exhibit even and odd parities under vertical
	reflection planes, for example, the ferromagnetic order $m_z$
	carries the $A_{2g}$ symmetry.
	$B_1$ and $B_2$ are analogous to the $d_{xy}$ and $d_{x^2-y^2}$
	symmetries, respectively, exhibiting opposite parities under the
	$\sigma_{x (y)}$ and $\sigma_{x^\prime (y^\prime)}$ operations.
	Symmetries of singlet channel gap functions are classified accordingly:
	The next-nearest neighbor (NNN) pairings
	are summarized in Tab.~III and the neighbor(NN)
	pairings in Tab.~IV in S. M. II \cite{supp_mat}.
	Different combinations of $f(\mathbf{k})$, $M$, and $\tau$ often
	lead to the equivalent symmetries, and in general, the existence
	of one can induce others in the same symmetry class.

	\begin{figure}[!htbp]
		\centering
		\includegraphics[width=0.9\linewidth]{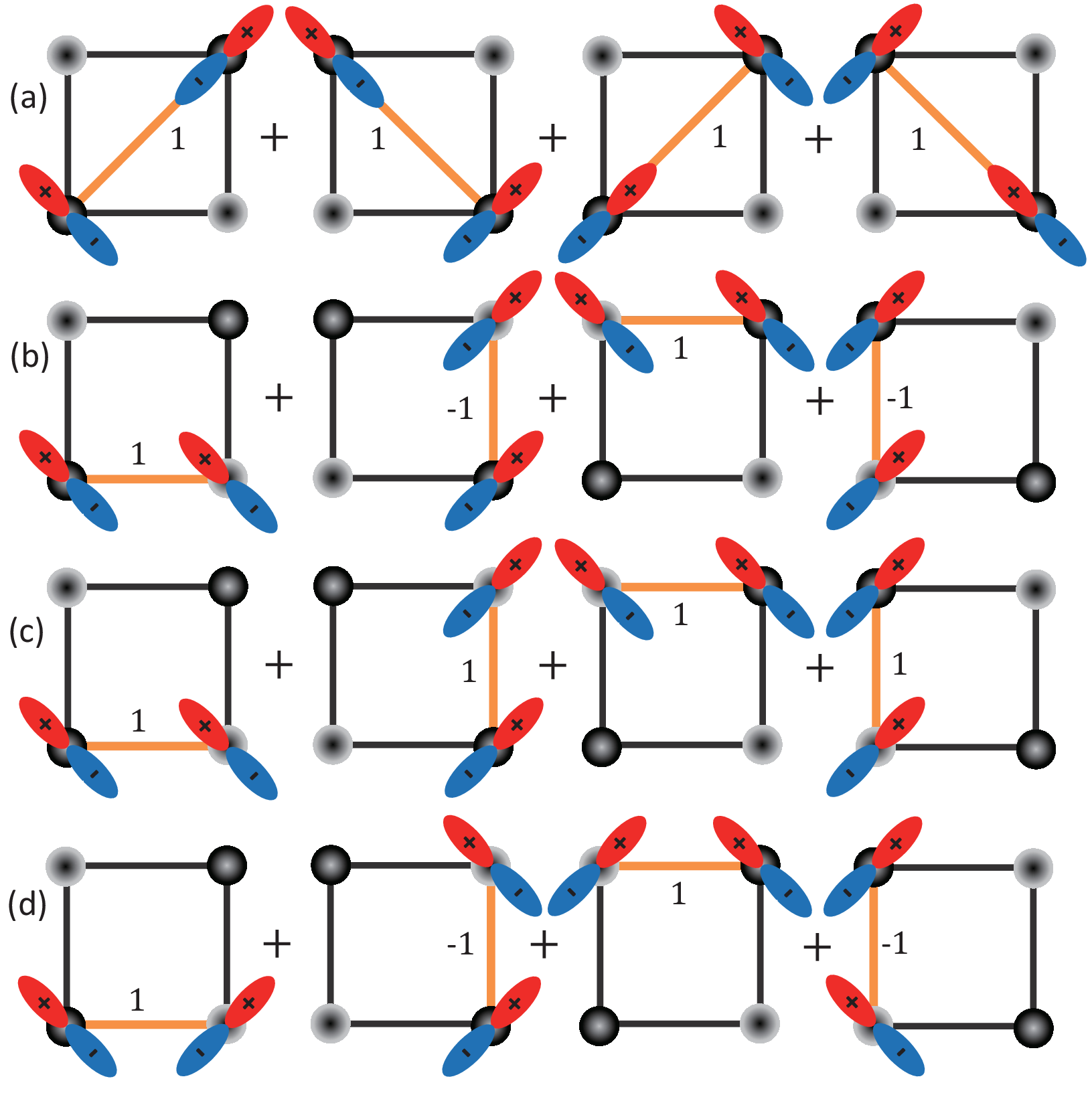}
		\caption{The real space orbital configurations for the singlet
			pairing on a square plaquette of the Fe cations.
			The $B_{1g}$ ($a$) and $B_{2g}$ ($b$) pairings cross the
			NNN and NN bonds, respectively, and their Fourier transforms
			give rise to the gap functions in Eq.~\eqref{eq:Bg}.
			The $A_{1g}$ ($c$) and $A_{2g}$ ($d$) pairings cross the NN bonds, and their Fourier transforms give rise to gap
			functions in Eq.~\eqref{eq:Ag}.
			Starting with the configuration of the first picture of each row,
			the subsequential configurations can be obtained by successively
			applying 4-fold rotations.
		}\label{fig:pair}
	\end{figure}
	
	Many orbital-dependent gap functions have been proposed in the literature \cite{nica_npjqm_2017,hu_arxiv_2018,nakayama_prb_2018,konig_arxiv_2018},
	and their importance have been analyzed in recent experiments \cite{sprau_science_2017,kostin_nat_mat_2018}.
	Due to the multi-orbital nature, the gap functions rigorously speaking
	cannot be intuitively represented by the partial-wave channels alone,
	i.e., the symmetry of the angular form factor $f(\mathbf{k})$.
	For example, consider the following two gap functions with even parity,
	\bea
	\Delta_1: \cos k_x\cos k_y\tau_0 \lambda_1, \ \ \,
	\Delta_2: (\cos k_x + \cos k_y)\tau_1 \lambda_3, \ \ \
	\label{eq:Bg}
	\eea
	which carry the $d_{xy}$ and $d_{x^2-y^2}$-like symmetries, or,
	more precisely, $B_{1g}$ and $B_{2g}$ symmetries, respectively,
	although their angular form factors are $s$-wave like.
	They involve the intra- and inter-orbital pairings between the
	$d_{x^\prime z}$ and $d_{y^\prime z}$-orbitals as shown in
	Fig. \ref{fig:pair} ($a$) and ($b$).
	Since $d_{xz}\to d_{yz}$ and $d_{yz}\to -d_{xz}$ under the $90^\circ$
	rotation, $\lambda_{1,3}$ transform analogously in the $d$-wave way.
	By examining their reflection symmetries, they belong to the $B_{1g}$
	and $B_{2g}$ symmetries.
	By similar analysis, the following gap functions, exhibit the $A_{1g}$
	and $A_{2g}$ symmetries, respectively, or, loosely speaking,
	the $s$-wave symmetry, in spite of their $d_{x^2-y^2}$ angular form factor:
	\bea
	\Delta_1: (\cos k_x - \cos k_y)\tau_1 \lambda_3, \ \ \,
	\Delta_2: (\cos k_x - \cos k_y)\tau_1 \lambda_1.  \ \ \, \ \ \,
	\label{eq:Ag}
	\eea
	Their orbital configurations are shown in Fig. \ref{fig:pair} ($c$)
	and ($d$).
	Furthermore, there can exist $p$-wave like pairing symmetry,
	or, the $E_g$-symmetry, in the singlet pairing channel,
	\bea
	\Delta_1: \cos(k_x+k_y) \tau_0 \lambda_4, \ \ \,
	\Delta_2: \cos(k_x-k_y) \tau_0 \lambda_6,
	\label{eq:Eg}
	\eea
	The former (latter) describes the pairing between the
	$d_{x^\prime z}$-orbital ($d_{y^\prime z}$) with the $d_{xy}$ one.

	
	The crystalline symmetries impose stringent constraints to
	the superconducting gap functions.
	According to Eq.~\eqref{eq:GL-mag-sc}, the direct product of the irreducible
	representations of $\Delta_1$ and $\Delta_2$ should contain that of
	$m_z$, i.e., $A_{2g}$.
	This yields the following possibilities of pairing symmetries:
	$B_{1g(u)}\pm i B_{2g(u)}$,
	$A_{1g(u)}\pm i A_{2g(u)}$ and $E_{g(u)}\pm i E_{g(u)}$.
	Examples of the above pairing symmetries with even parity
	are provided in Eqs. \eqref{eq:Bg}, \eqref{eq:Ag}, and \eqref{eq:Eg}.
	They carry the same symmetry under rotation, and hence, the experimental observables do not break the rotational symmetry. However, they mix different symmetries with respect to the vertical reflection plane, and such symmetries are also spontaneously breaking. For example, the magnetization $m_z$ is odd under such reflections.

	An important issue is that spin-orbit coupling is necessary to break
	the SU(2) symmetry such that the ferromagnetic order $m_z$ and
	superconducting orders $\Delta_{1,2}$ can couple, since the former
	and latter lie in the spin triplet and singlet channels, respectively.
	We employ a widely used three-band model for the topological band
	structure of FeTe$_{1-x}$Se$_x$ around the $\Gamma$-point, which consists of
	the $t_{2g}$-orbitals $d_{x^\prime z}$, $d_{y^\prime z}$ and $d_{xy}$ \cite{wang_prb_2015,xu_prl_2016}.
	Neglecting the small dispersion along the $z$-axis, the 3-band tight-binding Hamiltonian is expressed as \cite{daghofer2010},
	$H_0=\sum_k \psi^\dagger (k) {\cal H}_0(k) \psi(k)$
	where $\psi=\left\lbrack d_{x^\prime z,\uparrow}, d_{y^\prime z,\uparrow}, d_{xy,\uparrow}, d_{x^\prime z,\downarrow},
	d_{y^\prime z,\downarrow}, d_{xy,\downarrow} \right\rbrack^T$
	and the matrix kernel ${\cal H}_0(k)$ is given by,
	\bea
	\mathcal{H}_0 &= \mathcal{H}_{NNN} + \mathcal{H}_{NN}  + \mathcal{H}_{soc},
	\label{eq:ham0}
	\eea
	where $\mathcal{H}_{NNN}$ and $\mathcal{H}_{NN}$ represent the NNN and NN
	hoppings, respectively, with detailed forms presented in S. M. III \cite{supp_mat}.
	$\mathcal{H}_{soc} = \lambda_{soc} \tau_0 \vec{L}\cdot\vec{\sigma}$
	is the atomic spin-orbit coupling with $\lambda_{soc}$ the coupling strength
	and $\vec L$ representing the onsite orbital angular momentum projected
	to the $t_{2g}$-basis.
	Explicitly, $\vec{L}=\left( (\lambda_5-\lambda_7)/\sqrt{2},
	-(\lambda_5+\lambda_7)/\sqrt{2},-\lambda_2 \right)$.

	Based on the band structure Eq.~\eqref{eq:ham0}, the coupling coefficient
	$\gamma$ in Eq.~\eqref{eq:GL-mag-sc} can be evaluated as
	\bea
	\gamma &= \frac{1}{\beta}\sum_{\mathbf{k},\omega_n} f_1^{l_1} (\mathbf{k})f_2^{l_2}(\mathbf{k})
	\text{Tr}\left\lbrack G_h M_1^{l_1} G_e \sigma_z G_e M_2^{l_2} \right\rbrack,
	\label{eq-gamma}
	\eea
	where $\beta=1/(k_B T)$ is the inverse of temperature;
	$f^{l_i}_i$ and $M_i^{l_i}$ with $i=1,2$ are the angular form factors
	and orbital pairing matrix kernels of $\Delta_{1,2}$, respectively;
	$G_e(\mathbf{k},i\omega_n) = (i\omega_n - \mathcal{H}_0(\mathbf{k}))^{-1}$
	is the Matsubara Green's function with  $\omega_n=(2n+1)\pi/\beta$,
	and $G_h(\mathbf{k},i\omega_n)=G_e^* (-\mathbf{k},-i\omega_n)$.
	As shown in S. M. III \cite{supp_mat}, the hole-like Fermi pockets around the $\Gamma$-point
	are mainly from the bonding states between two Fe-sublattices,
	{\it i.e.}, they are approximately eigenstates of $\tau_1$ with
	the eigenvalue of 1.
	Hence, only tracing over the spin and orbital channels are needed,
	and only these gap functions characterized by $\tau_{0,1}$ are considered.
	Gap functions with $\tau_{2,3}$ are pairing between bonding
	and anti-bonding states between two-sublattices, which
	will be neglected below.
	
	
	Next we present the examples of gap functions $\Delta_{1,2}$ leading to
	the spontaneous magnetization $m_z$.
	We begin with the cases of $B_{1g(u)}\pm i B_{2g(u)}$.
	For parity even, i.e., $B_{1g}\pm i B_{2g}$, we take
	$\Delta_{1,2}$ in the form of Eq.~\eqref{eq:Bg}.
	As shown in S. M. IV \cite{supp_mat}, after further reducing the band Hamiltonian
	Eq.~\eqref{eq:ham0} to a two-band model only based on $d_{x^\prime(y^\prime)z}$,
	Eq.~\eqref{eq-gamma} yields an analytic expression as
	\bea
	\gamma \approx  -\frac{7\zeta(3)}{4\pi^3} \frac{\lambda_{soc} N_0}{(k_B T_c)^2},
	\eea
	with $N_0$ the density of states at the Fermi surface.
	Since $m_z$ is induced by the TR breaking pairings via spin-orbit coupling,
	the coupling coefficient $\gamma$ is proportional to the spin-orbit coupling strength.
	A calculation based on the 3-band Hamiltonian is also performed numerically,
	which yields consistent results (see S. M. V \cite{supp_mat}).
	The parity odd case, i.e., $B_{1u}\pm i B_{2u}$, is also numerically
	checked to yield a nonzero $\gamma$, for example, with $\Delta_1:
	\tau_0(\sin(k_x+k_y)\lambda_7 + \sin(k_x-k_y)\lambda_5)$ and
	$\Delta_2: \tau_0(\sin(k_x+k_y)\lambda_5 - \sin(k_x-k_y)\lambda_7)$.
	The above two cases 
	also break the mirror symmetries of $\sigma_{x (y)}$ and
	$\sigma_{x^\prime (y^\prime)}$ spontaneously.
	They are topologically non-trivial belonging to the C-class supporting
	the chiral Majorana edge modes \cite{hasan2010,qi2011,bansil2016,chiu2016}.
	We have also studied both cases of $A_{1g(u)}\pm i A_{2g(u)}$, which
	also yield non-zero $\gamma$'s as shown in S. M. V \cite{supp_mat}.
	The nodal pairing gap functions presented in Eq.~\eqref{eq:Ag} are
	used for the even parity case, and the nodeless pairing with
	$\Delta_1: \tau_0 (\sin(k_x+ k_y) \lambda_5 + \sin(k_x-k_y)\lambda_7)$
	and $\Delta_2: \tau_0(\sin(k_x+k_y)\lambda_7 - \sin(k_x-k_y)\lambda_5)$
	are used for the odd parity case.
	For the case of $E_{g(u)}\pm i E_{g(u)}$, we take the gap functions
	presented in Eq.~\eqref{eq:Eg} as an example.
	Since $E_{g(u)}\otimes E_{g(u)} = A_{1g} \oplus A_{2g} \oplus B_{1g} \oplus
	B_{2g}$, it also yields a nonzero $\gamma$ as calculated in S. M. IV \cite{supp_mat}, and consequently induces magnetization.
	
	In strongly correlated superconductors, there exist strong superconducting
	phase fluctuations in the normal state close to $T_c$ \cite{emery1995}.
	In this case, the phases $\varphi_1$ and $\varphi_2$ of gap functions
	are disordered such that $\avg{\Delta_{1,2}}=0$, but $|\Delta_{1,2}|$
	remains finite.
	The $\beta^\prime$ term in Eq.~\eqref{eq:GL} can still pin the relative
	phase $\Delta \varphi=\pm \frac{\pi}{2}$.
	This transition breaks TR symmetry and its critical temperature
	$T^\prime>T_c$.
	We still expect a weaker but still finite $m_z$ in the
	temperature window between $T_c$ and $T^\prime$.

	The bulk spin magnetization induced by TR breaking pairing qualitatively
	explains the gap opening of the surface Dirac cone observed in the FeSe$_{0.3}$Te$_{0.7}$ superconductor \cite{johson_expt}.
	We next briefly discuss the issue of the possible magnetic field generated
	by the bulk magnetization.
	Its upper bound is estimated to be $15$ Gauss (See S. M. VI \cite{supp_mat}), which is
	still smaller than the lower critical magnetic field of the FeSe$_{1-x}$Te$_x$
	superconductors ($H_{c1}\sim30$ Gauss)\cite{yadav2009,abdel2013}.
	The actual field due to spin magnetization should be much smaller
	than this bound, hence, it can be offset by the orbital
	magnetization such that the total magnetic field remains zero
	in the Meissner state \cite{shopova2005,chirolli2017}.
	This is consistent with the observation that neutron spectroscopy does not detect a bulk magnetic field \cite{xu2018}.
	A very weak but finite internal magnetic field around $0.15$ Gauss is
	detected in the FeSe superconductor by the muon spin rotation measurement ($\mu$SR)\cite{watashige2015,matsuura2019}, which may arise
	from the imperfect screening due to impurities and domains.

	\textit{Discussion and Conclusion--.}
	In this article, we have studied how the TR symmetry breaking superconducting
	states can gap out the topological surface modes in the iron-chalcogenide superconductors.
	Spin-orbit coupling is necessary to break the SU(2) symmetry in the
	spin channel, such that it bridges the magnetic ordering and the
	TR breaking pairing states.
	Three classes of gap function symmetries can lead to such an effect
	based on group theory analyses: $A_{1g(u)}+iA_{2g(u)}$,
	$B_{1g(u)}+iB_{2g(u)}$, $E_{g(u)}+iE_{g(u)}$.
	In strongly correlated superconductors, the superconducting
	phase fluctuations can also lock their relative phase
	at $\pm\frac{\pi}{2}$ breaking TR symmetry in the normal state.
	This work builds connections between novel pairing symmetries
	in iron-based superconductors and their topological band structures.
	In particular, it is helpful for understanding the superconductivity induced gap
	opening for the topological surface state recently discovered
	in the FeTe$_{0.7}$Se$_{0.3}$ \cite{johson_expt}.
	
	\textit{Note added.}
	During the review process, we become aware of a recent work by T.~Kawakami \cite{kawakami_prb_2019}, where they mainly discuss the topological odd-parity superconductivity in iron-based superconductors.

	{\it Acknowledgments--.}
	L. H., P. D. J., and C. W. acknowledge helpful discussions with
	J. Tranquada, G. D. Gu.
	L. H. and C. W. are supported by AFOSR FA9550-14-1-0168.
	P. D. J. is supported by
	the Office of Science, U.S. Department of Energy under contract number DE-SC001274.

	\bibliographystyle{apsrev4-1}
	\bibliography{ref}

\clearpage
\appendix
\section{P4/nmm non-symmorphic group}
Here we review the $P4/nmm$ group for the FeSe crystal.
If the Fe site is chosen as the origin, there exists
the $D_{2d}$ point group symmetry centering at the Fe site, consisting of operations of $\left\{E, S_4, C^\prime_{2}(z), S_4^3, C_{2}(x), C_{2}(y), \sigma_{x'}, \sigma_{y'} \right\}$, where $C^\prime_2(z)$ is denoted
to distinguish the $C_2(z)$ centering at Se/Te site below.
Each unit cell contains two Fe and Se atoms exhibiting a square lattice
with the lattice constant $\sqrt{2}a$ where $a$ measures the nearest neighbor
Fe-Fe bond in the $x$-$y$ plane.
It is also convenient to choose the Se/Te site as the origin, then the
corresponding point group is $C_{4v}=\left\{E,C_{4}(z), C_2(z), C^3_4(z),
\sigma_{x}, \sigma_{y}, \sigma_{x'}, \sigma_{y'}\right\}$,
which is used in this work to specify the irreducible representations
of gap functions.
The configurations of the reflection planes of $\sigma_x,\sigma_y,
\sigma_{x^\prime}$ and $\sigma_{y^\prime}$ are given in Fig. 2
in the main text.

The $P4/nmm$ space group of the FeSe crystal can be decomposed into the
cosets denoted as $g T$, where $g$ is a symmetry operation;
$T$ is the lattice translation group consisting of translations of $\sum_{i=1,2,3}l_i\vec{a}_i$ with integers $l_{1,2,3}$,
and the lattice vectors are defined as $\vec{a}_1=(a,a,0)$,
$\vec{a}_2=(a,-a,0)$ and $\vec{a}_3=(0,0,c)$ with $c$
the length of Fe-Fe bond along the $z$-direction.
There are 8 cosets generated by $g \in C_{4v}$ point group:
\bea
&&T, \ \ \  C_4^1(z)T, \ \ \ C_2(z)T, \ \ \ C_4^3(z)T,  \\
&&\sigma_{x} T, \ \ \  \sigma_y T, \ \ \  \sigma_{x^\prime} T,
\ \ \ \sigma_{y^\prime} T, \nonumber
\eea
where $C_4^{1,3}(z)$ and $C_2(z)$ are rotations
around the $z$-axis passing the Se atom, and $\sigma_{x(y)}$ and $\sigma_{x^\prime(y^\prime)}$ are reflections
with respect to the vertical planes passing the Se atoms as
shown in Fig. 2 in the main text.
The rest cosets are constructed by applying the inversion operation
$I$ with respect
to the Fe-Fe bond centers to the previous 8 ones, which are
\bea
&&IT, \ \ \ S_4^3 T, \ \ \ g(\sigma_h,\tau) T, \ \ \ S_4^1 T,  \\
&& C_2(y) T, ~~ C_2(x) T, ~~ g(C_2(y^{\prime\prime}),\vec \tau_y)T,
~~
g(C_2(x^{\prime\prime}),\vec \tau_x)T, \nn
\eea
where $S_4$ and $S_4^3$ are the rotary reflection centering around
the Fe cation;
$C_2(x)$ and $C_{2}(y)$ are the 2-fold rotations around
the nearest Fe-Fe bonds.
There are three class of non-symmorphic operations:
$g(\sigma_h, \vec \tau)$ is the glide reflection
with $\sigma_h$ the reflection with respect to the $xy$-plane
and $\vec \tau=a(1,0,0)$;
$g(C_2(x^{\prime\prime}),\vec \tau_x)$ and
$g(C_2(y^{\prime\prime}), \vec\tau_y)$
are screw rotations with their axes $x^{\prime\prime}$
and $y^{\prime\prime}$ along the $45^\circ$ and $135^\circ$ degrees
passing the Fe-Fe bond centers;
$\vec \tau_x=\frac{1}{2}(-a, a,0)$ and $\vec \tau_y=\frac{1}{2}(a,a,0)$.

The character table of $C_{4v}$ is given in Table.~\ref{tab-c4v}, where
there are four one-dimensional irreducible representations $A_1$, $A_2$,
$B_1$, $B_2$ and one two-dimensional representation $E$.
Taking the inversion symmetry of $P4/nmm$ into account, the total
representations are: $A_{1g/1u}$, $A_{2g/2u}$, $B_{1g/2u}$,
$B_{2g/2u}$ and $E_{g/u}$, where $g$ and $u$ mean even and odd under
the inversion symmetry, respectively.

\begin{table}[!htbp]
	\begin{ruledtabular}
		\begin{tabular}{c|c|c|c|c|c}
			$C_{4v}$ & $E$   & $2C_{4}(z)$ & $C_2(z)$ & $2\sigma_{x'}$ & $2\sigma_{x}$    \\ \cline{1-6}
			$A_1$    & $1$  & $1$        & $1$     & $1$            & $1$ \\ \hline
			$A_2$    & $1$  & $1$        & $1$     & $-1$            & $-1$ \\ \hline
			$B_1$    & $1$  & $-1$        & $1$     & $1$            & $-1$ \\ \hline
			$B_2$    & $1$  & $-1$        & $1$     & $-1$            & $1$ \\ \hline
			$E$      & $2$  & $0$         & $-2$     & $0$             & $0$ \\
\end{tabular}
\end{ruledtabular}
\caption{The character table for the point group $C_{4v}$.}
\label{tab-c4v}
\end{table}

\section{The gap function symmetry}
\begin{figure}[!htbp]
	\centering
	\includegraphics[width=0.5\linewidth]{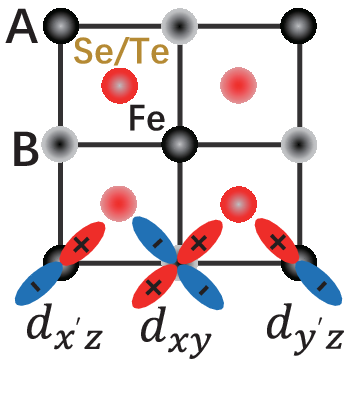}
	\vspace{-5mm}
	\caption{The orbital configurations of $d_{x^\prime z}$, $d_{y^\prime,z}$
and $d_{xy}$ on the Fe cations.
	}
	\label{fig-sup-dorbitals}
\end{figure}

According to the crystal symmetry, the $d_{x'z},d_{y'z},d_{xy}$-orbitals
on the A and B sublattices of Fe are constructed shown in Fig. \ref{fig-sup-dorbitals}, where the $d_{x^\prime z}$ and $d_{y^\prime z}$
are along the diagonal Fe-Fe lines.
For the three dimensional orbital space, we define the $3\times 3$ Gell'mann
matrices according to the basis of $(d_{x'z},d_{y'z},d_{xy})^T$ as
\begin{align}
 \lambda_1 &= \left(\begin{array}{ccc} 0 & 1  & 0  \\  1 & 0  & 0  \\  0 & 0  & 0  \end{array}\right),\quad
 \lambda_2  = \left(\begin{array}{ccc} 0 & -i & 0  \\  i & 0  & 0  \\  0 & 0  & 0  \end{array}\right), \\
 \lambda_3 &= \left(\begin{array}{ccc} 1 & 0  & 0  \\  0 & -1 & 0  \\  0 & 0  & 0  \end{array}\right),\quad
 \lambda_4  = \left(\begin{array}{ccc} 0 & 0  & 1  \\  0 & 0  & 0  \\  1 & 0  & 0  \end{array}\right), \\
 \lambda_5 &= \left(\begin{array}{ccc} 0 & 0  & -i \\  0 & 0  & 0  \\  i & 0  & 0  \end{array}\right),\quad
 \lambda_6  = \left(\begin{array}{ccc} 0 & 0  & 0  \\  0 & 0  & 1  \\  0 & 1  & 0  \end{array}\right), \\
 \lambda_7 &= \left(\begin{array}{ccc} 0 & 0  & 0  \\  0 & 0  & -i \\  0 & i  & 0  \end{array}\right),\quad
 \lambda_8  = \frac{1}{\sqrt{3}}\left(\begin{array}{ccc} 1 & 0  & 0  \\  0 & 1  & 0  \\  0 & 0  & -2 \end{array}\right).
\end{align}
In addition, $\lambda_0$ is the $3\times 3$ identity matrix.
We also denote the Pauli matrices in the channel of the Fe A/B sublattice
as $\tau_{1,2,3}$ and $\tau_0$  the $2\times 2$ identity matrix.
Similarly, the Pauli matrices in the spin channel is denoted as
$\sigma_{x,y,z}$, and $\sigma_0$ is also the $2\times 2$ identity matrix.

The classification for matrices in the sublattice channel ($\tau$),
the orbital channel ($\lambda$), and the spin channel ($\sigma$)
according to the $C_{4v}$ and inversion symmetries are summarized
in the following Table.~\ref{tab-matrice-representation}.

\begin{table}[!htbp]
	\begin{ruledtabular}
		\begin{tabular}{c|c|c|c} Irreps. & A/B Sub-lattice & Orbital & Spin \\ \cline{1-4}
				$A_{1g}$ & $\tau_0$, $\tau_1$ & $\lambda_0$, $\lambda_8$ & $\sigma_0$  \\ \hline
				$A_{2g}$ &   /  & $\lambda_2$  & $\sigma_z$ \\ \hline
				$B_{1g}$ &   /  & $\lambda_1$  &  /   \\ \hline
				$B_{2g}$ &   /  & $\lambda_3$  &  /   \\ \hline
				$B_{2u}$ & $\tau_2$, $\tau_3$ &  /   & / \\ \hline
				$E_{g}$  &   /  & $\{\lambda_4,\lambda_6\}$, $\{\lambda_5,\lambda_7\}$ & $ \{\sigma_x,\sigma_y\}$ \\
		\end{tabular}
\end{ruledtabular}
\caption{Representation for matrices in different channels: $\tau$'s
in the channel of  the iron A/B sublattice, $\lambda$'s in the
orbital channel, and $\sigma$'s in the spin channel.
For the $E_g$ representation, the pair of the $\lambda$-matrices
inside the braces form a two-dimensional basis.
}\label{tab-matrice-representation}
\end{table}

\begin{table}[!htbp]
\begin{ruledtabular}
\begin{tabular}{c|c|c} & $u$ & $g$  \\ \cline{1-3}
$A_1$ & $\sin k_x \sin k_y \tau_3 \lambda_{0,8} $ & $\cos k_x \cos k_y \tau_0\lambda_{0,8} $\\
      & $\cos k_x \cos k_y\tau_3 \lambda_{3} $   & $ \sin k_x \sin k_y \tau_0\lambda_{3}$ \\
      & $i\tau_0( \sin k_+ \lambda_5 + \sin k_-\lambda_7 )$ & $i\tau_3(\sin k_+ \lambda_5+\sin k_- \lambda_7)$ \\ \hline
$A_2$ & $\cos k_x \cos k_y \tau_3\lambda_1 $  &  $\sin k_x \sin k_y\tau_0\lambda_1 $ \\
      & $i \tau_0(\sin k_+ \lambda_7- \sin k_- \lambda_5)$ & $i\tau_3(\sin k_+ \lambda_7-\sin k_- \lambda_5)$\\ \hline
$B_1$ & $ \sin k_x \sin k_y \tau_3\lambda_1 $ & $ \cos k_x \cos k_y \tau_0\lambda_1 $ \\
      & $i\tau_0(\sin k_+ \lambda_7+ \sin k_-\lambda_5)$ & $i\tau_3(\sin k_+ \lambda_7+ \sin k_- \lambda_5)$\\ \hline
$B_2$ & $ \cos k_x \cos k_y \tau_3\lambda_{0,8}$ & $\sin k_x \sin k_y \tau_0\lambda_{0,8}$ \\
      & $ \sin k_x \sin k_y \tau_3\lambda_3 $    & $ \cos k_x \cos k_y \tau_0\lambda_{3} $\\
      & $i\tau_0( \sin k_+ \lambda_5 - \sin k_- \lambda_7 )$ & $i\tau_3( \sin k_+ \lambda_5- \sin k_- \lambda_7)$ \\ \hline
$E$   & $( \sin k_+, ~ \sin k_-)i\tau_0\lambda_2$ & $( \sin k_+, ~ \sin k_-)i\tau_3\lambda_2$ \\
      & $\tau_3( \cos k_+ \lambda_4, ~ \cos k_-\lambda_6)$ & $\tau_0(\cos k_+ \lambda_4, ~\cos k_-\lambda_6)$ \\
      & $\tau_3( \cos k_+ \lambda_6, ~\cos k_-\lambda_4)$ & $\tau_0( \cos k_+\lambda_6, ~ \cos k_-\lambda_4)$
\end{tabular}
\end{ruledtabular}
\caption{Classifications of the singlet superconducting order parameters
for the NNN pairings according to the symmetries of $C_{4v}$ and
inversion.
$k_\pm=k_x\pm k_y$;
$u$ and $g$ represent the odd and even parities for the
inversion with respect to the Fe-Fe bond centers, respectively.
Each line for the 2D $E$-representation contains a pair of
degenerate gap functions.
}
\label{tab-pairing-nnn}
\end{table}

The pairing gap functions in the spin singlet channels and their
symmetry properties under the $C_{4v}$ group and the inversion
operations are systematically analyzed.
Symmetries for pairings across the NNN Fe-Fe bonds are summarized
in Tab. III, and those for pairings along the NN Fe-Fe bonds
are summarized in Tab. IV.

\begin{table}[!htbp]
\begin{ruledtabular}
\begin{tabular}{c|c|c}  & $u $ & $g$  \\ \cline{1-3}
$A_1$ & $(\cos k_x -\cos k_y)\tau_2\lambda_2$ & $(\cos k_x+\cos k_y)\tau_1\lambda_{0,8}$\\
      &    & $(\cos k_x -\cos k_y)\tau_1\lambda_{1}$ \\
      & $i \tau_1(\sin k_x \lambda^{57}_+ + \sin k_y \lambda_-^{57} )$ & $i\tau_2(\sin k_x \lambda_-^{46} + \sin k_y \lambda_+^{46} )$ \\ \hline
$A_2$ & $ $ &  $(\cos k_x-\cos k_y)\tau_1\lambda_{3}$ \\
      & $i\tau_1(\sin k_x \lambda_-^{57} + \sin k_y \lambda_+^{57} )$ & $i\tau_2(\sin k_x \lambda_+^{46} + \sin k_y \lambda_-^{46})$\\ \hline
$B_1$ & $(\cos k_x +\cos k_y)\tau_2\lambda_2$ & $(\cos k_x -\cos k_y )\tau_1\lambda_{0,8}$ \\
      & $ $ & $(\cos k_x +\cos k_y)\tau_1\lambda_{1}$\\
      & $i\tau_1(\sin k_x \lambda_+^{57} - \sin k_y \lambda_-^{57} )$ & $\tau_2(\sin k_x \lambda_-^{46} - \sin k_y \lambda_+^{46} )$ \\\hline
$B_2$ & $ $ & $(\cos k_x +\cos k_y )\tau_1\lambda_{3} $ \\
      & $i \tau_1(\sin k_x \lambda_-^{57} - \sin k_y \lambda_+^{57} )$    & $i\tau_2(\sin k_x \lambda_+^{46} - \sin k_y \lambda_-^{46} )$ \\ \hline
$E$   & $(\sin k_x, ~ \sin k_y)i\tau_1\lambda_2 $ & $(\sin k_x, ~ \sin k_y)i\tau_2\lambda_{0,8} $ \\
      &   & $(\sin k_x, ~\sin k_y)i\tau_2\lambda_{1,3} $ \\
      & $i \tau_2(\cos k_x \lambda_-^{57}, ~\cos k_y \lambda_+^{57})$ & $\tau_2(\cos k_x \lambda_-^{46}, ~ \cos k_y \lambda_+^{46})$ \\
      & $i \tau_2( \cos k_x \lambda_+^{57}, ~\cos k_y \lambda_-^{57})$ & $i \tau_2(\cos k_x \lambda_+^{46},~ \cos k_y \lambda_-^{46})$
\end{tabular}
\end{ruledtabular}
\caption{Classifications of the singlet superconducting order parameters
for the NN pairings.
$\lambda_\pm^{46} =\lambda_4\pm\lambda_6$ and $\lambda_\pm^{57}=\lambda_5\pm\lambda_7$.
	}\label{tab-pairing-nn}
\end{table}

\section{Three-orbital model}
\begin{figure*}[!htbp]
	\centering
	\includegraphics[width=0.8	\linewidth]{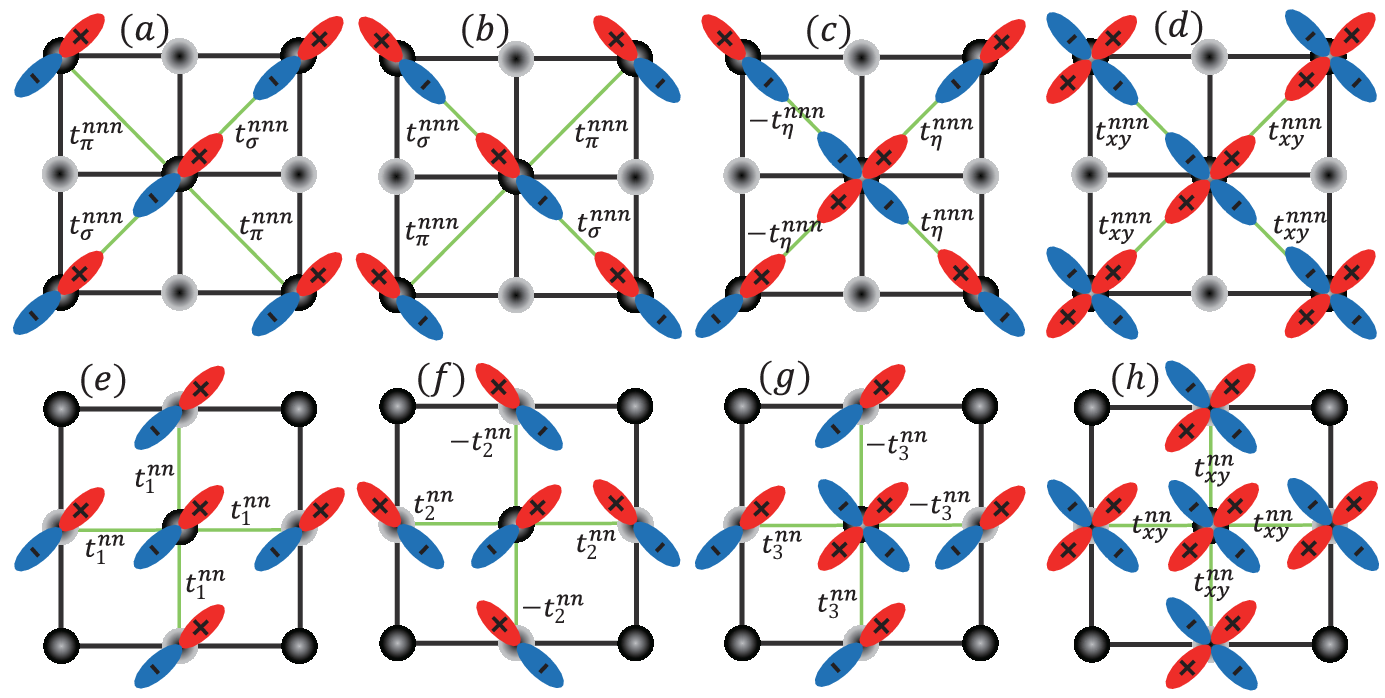}
	\vspace{-1mm}
	\caption{The eight independent hoping integrals. The NNN hoping integrals $t_{\sigma,\pi,xy,\eta}^{NNN}$ are shown in (a-d), and the NN hoping integrals $t_{1,2,3,xy}^{NN}$ are shown in (e-h).
	}
	\label{fig-sup-hoppings}
\end{figure*}

The corresponding tight-binding model can be constructed below,
\bea\label{supp-eq-ham0}
  H_0 &= \sum_{\mathbf{k}} \psi^\dagger(\mathbf{k}) \mathcal{H}_0(\mathbf{k}) \psi(\mathbf{k}),
\eea
where $\psi(\mathbf{k}) =\lbrack d_{A,x'z,\uparrow}, d_{A,y'z,\uparrow}, d_{A,xy,\uparrow}, d_{B,x'z,\uparrow}, d_{B,y'z,\uparrow},\\ d_{B,xy,\uparrow},  d_{A,x'z,\downarrow}, d_{A,y'z,\downarrow}, d_{A,xy,\downarrow}, d_{B,x'z,\downarrow}, d_{B,y'z,\downarrow}, d_{B,xy,\downarrow} \\ \rbrack^T$,
and the matrix kernel is defined as
\bea
 \mathcal{H}_0 &= \tau_0 \mathcal{H}_{NNN} + \tau_1 \mathcal{H}_{NN} + \mathcal{H}_{soc}.
 \label{eq:matrix}
\eea
The $\mathcal{H}_{NNN}$ term describes the next-nearest neighbor (NNN)
hopping Hamiltonian,
\bea
\mathcal{H}_{NNN} &=& 2(t_{\sigma}^{nnn}+t_\pi^{nnn}) \cos k_x \cos k_y \left(2\lambda_0 + \sqrt{3}\lambda_8 \right)/3 \nonumber \\
     &+& (4t_{xy}^{nnn} \cos k_x \cos k_y +\Delta_{xy}) \left(\lambda_0 - \sqrt{3}\lambda_8 \right)/3 \nonumber \\
     &+& 2(t_\pi^{nnn}-t_\sigma^{nnn}) \sin k_x \sin k_y \lambda_3 \\
     &+& 2t_\eta^{nnn}[\sin(k_x+k_y)\lambda_5 + \sin(k_x-k_y)]\lambda_7, \nn
\eea
where $\Delta_{xy}$ is the the on-site energy difference between the $d_{xy}$-orbital and the two degenerated $d_{x'z/y'z}$;
the hopping integrals $t^{nnn}_{\sigma,\pi,\eta,xy}$ here describe
the bonding along the NNN Fe-Fe bond (see Fig.~\ref{fig-sup-hoppings}(a-d)).
$t^{nnn}_{\sigma}$ and $t^{nnn}_{\pi}$ are the $\sigma$ and $\pi$-bonding
strengths of the $d_{x^\prime(y^\prime),z}$ orbitals;
$t_{xy}^{nnn}$ is the bonding strength between two $d_{xy}$-orbitals;
$t_\eta^{nnn}$ is the bonding between $d_{x^\prime(y^\prime)z}$ and
$d_{xy}$-orbitals.

The $\mathcal{H}_{NN}$ term represents the nearest neighbor (NN)
hopping Hamiltonian,
\begin{align}
 \mathcal{H}_{NN} &= 2t_1^{nn} (\cos k_x + \cos k_y) \left(2\lambda_0 + \sqrt{3}\lambda_8 \right)/3 \nonumber \\
     &+ 2t_{xy}^{nn} (\cos k_x + \cos k_y) \left(\lambda_0 - \sqrt{3}\lambda_8 \right)/3 \nonumber \\
     &+ 2t_2^{nn}(\cos k_x - \cos k_y) \lambda_1  \\
     &- 2t_3^{nn} [(\sin k_x+\sin k_y)\lambda_5
     + (\sin k_x - \sin k_y)\lambda_7 ], \nonumber
\end{align}
where $t^{nn}_{1,2,3,xy}$ describe the bonding along the NN
Fe-Fe bond (see Fig.~\ref{fig-sup-hoppings}(e-h)).
$t_1^{nn}$ and $t_2^{nn}$ describe the intra- and inter-orbital
bonding between $d_{x^\prime z}$ and $d_{y^\prime z}$-orbitals;
$t^{nn}_{xy}$ is the bonding between $d_{xy}$-orbitals;
$t^{nn}_3$ is the bonding between $d_{xy}$ and $d_{x^\prime(y^\prime) z}$
orbitals.

The last term $\mathcal{H}_{soc}$ is the spin-orbit coupling Hamiltonian,
\begin{align}
  \mathcal{H}_{soc} = \lambda_{soc} \tau_0 \left(\vec{L}\cdot\vec{\sigma}\right).
\end{align}
where $\vec{L}=\left( (\lambda_5-\lambda_7)/\sqrt{2},
-(\lambda_5+\lambda_7)/\sqrt{2},-\lambda_2 \right)$.

\begin{figure}[!htbp]
	\centering
	\includegraphics[width=0.8\linewidth]{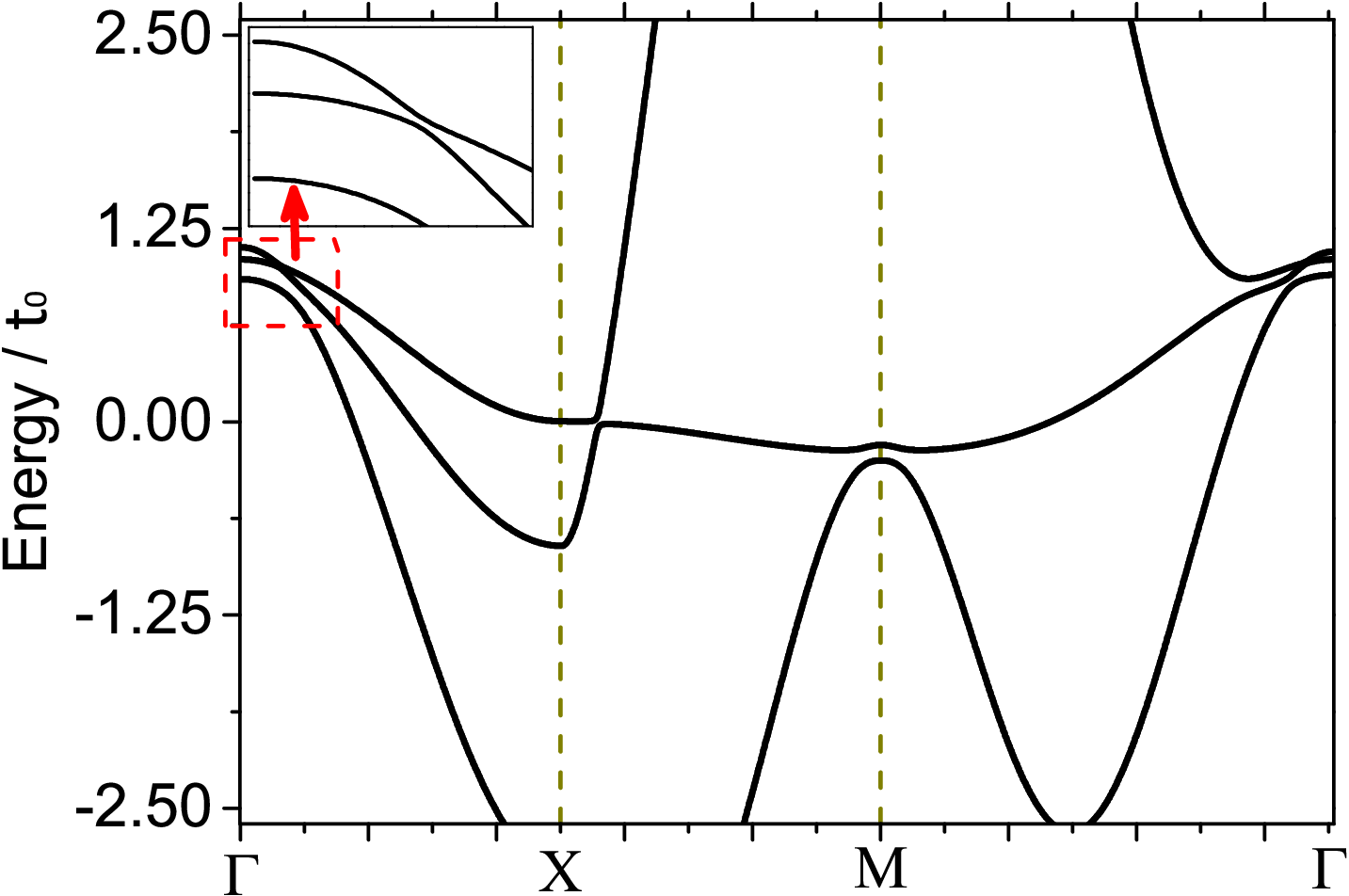}
	\caption{Band structure along $\Gamma-X-M-\Gamma$ lines in the unit of $t_0$. Parameters are used as: $t_\sigma^{nnn}=0.15t_0$, $t_\pi^{nnn}=0.05t_0$, $t_{xy}^{nnn}=t_0$,
		$t_\eta^{nnn}=t_0/\sqrt{2}$,
$t_1^{nn}=0.15t_0$, $t_{xy}^{nn}=-t_0$, $t_2^{nn}=-0.1t_0$, $t_3^{nn}=t_0/\sqrt{2}$, $\Delta_{xy}=1.1t_0$ and $\lambda_{soc}=0.05t_0$. All the states are two-fold degenerated because of time-reversal symmetry and inversion symmetry.
	}
	\label{fig-supp-bands}
\end{figure}

Since $\tau_1$ is conserved in Eq.\eqref{eq:matrix}, we can label
the eigenstates of ${\cal H}_0$ by $\tau_1$'s eigenvalues of $\pm 1$.
There are mainly three hole-like pockets around the $\Gamma$-point,
and all of them carry the eigenvalue of 1 of $\tau_1$.
Projecting ${\cal H}_0$ into this sector, we arrive at a $6\times 6$
Hamiltonian matrix as
\begin{align}\label{supp-eq-ham0-3orb}
\mathcal{H}_0 = \mathcal{H}_{NNN} + \mathcal{H}_{NN}.
\end{align}
The corresponding band structure is calculated along the high symmetry
lines $\Gamma$-$X$-$M$-$\Gamma$ as shown in Fig. \ref{fig-supp-bands}
with the parameters given in the figure caption.
Each band is doubly degenerate due to the TR symmetry and
inversion symmetry.

\section{The reduced two-orbital model}

Due to the relatively large value of $\Delta_{xy}$, we can project out
the $d_{xy}$-orbital and arrive a reduced two-orbital model with
only $d_{x^\prime z}$ and $d_{y^\prime z}$-orbitals, whose matrix
kernel read
\bea
\bar{ \cal {H}}_0 &= \tau_0 \bar{\cal {H}}_{NNN}
+ \tau_1 \bar{\cal {H}}_{NN},
\eea
with
\bea
\bar{\cal {H}}_{NNN} &=& 2\bar{t}_1 \cos k_x \cos k_y \bar{\lambda}_0 + 2\bar{t}_2 \sin k_x \sin k_y \bar{\lambda}_3, \nn \\
\bar{\cal{H}}_{NN} &=& 2\bar{t}_5(\cos k_x + \cos k_y) \bar{\lambda}_0  + 2\bar{t}_7(\cos k_x - \cos k_y) \bar{\lambda}_1,
\nn
\eea
where $\bar{t}^{nnn}_{\sigma,\pi}$ are hopping elements for NNN Fe-Fe bonding;
$\bar{t}^{nn}_{1,2}$ are those for the NN Fe-Fe bonding; the $\bar{\lambda}_{1,2,3}$ are Pauli matrices defined in the orbital channel ($d_{x'z}$ and $d_{y'z}$), and $\bar{\lambda}_0$ is the $2\times 2$
identity matrix.
Then there will be two hole-like pockets around the $\Gamma$-point and
all the states on the Fermi surface are eigenstates of $\tau_1$ with
the eigenvalues $+1$.
Within this sector, the $\bar{ \cal {H}}_0$ can be further simplified as
\bea
 \bar{\cal{H}}_0 = \bar{\cal{H}}_{NNN} + \bar{\cal{H}}_{NN}
 + \bar{\cal{H}}_{soc},
\eea
where $\bar{\cal{H}}_{soc} = \lambda_{soc} \bar{\lambda}_2 \sigma_z$.
Then the Green's function can be analytically solved as,
\begin{align}
		G_e &= \frac{\mathcal{P}_+}{-i\omega_n + \epsilon_+} +  \frac{\mathcal{P}_-}{-i\omega_n + \epsilon_-}, \nn \\
	\sigma_y G_h \sigma_y &= \frac{\mathcal{P}_+}{-i\omega_n - \epsilon_+} +  \frac{\mathcal{P}_-}{-i\omega_n - \epsilon_-},
\end{align}
where $\beta=1/T$ ($T$ is temperature) and the fermion Matsubara frequency $\omega_n=(2n+1)\pi/\beta$. The projection operators are defined as
\begin{align}
\begin{split}
\mathcal{P}_{\pm} &= \frac{1}{2}\Big{\{} 1 \pm \big{(}  2\bar{t}_2 \sin k_x \sin k_y \bar{\lambda}_3 \\
 &+ 2\bar{t}_7(\cos k_x - \cos k_y) \bar{\lambda}_1  + \lambda_{soc} \bar{\lambda}_2 \sigma_z \big{)}/E_k \Big{\}},
\end{split}
\end{align}
with
\bea
E_k &=& \sqrt{4\bar{t}_2^2 \sin^2 k_x \sin^2 k_y + 4\bar{t}_7^2(\cos k_x - \cos k_y)^2 +\lambda_{soc}^2 }, \nn \\
\epsilon_\pm &=&\epsilon_0(\mathbf{k}) \pm E_k,
\eea
where $\epsilon_0 = 2\bar{t}_1 \cos k_x \cos k_y + 2\bar{t}_5(\cos k_x + \cos k_y)$.
This simplification helps to calculate the Feymann diagram analytically
discussed in the main text.



\begin{figure}[!htbp]
\centering
\includegraphics[width=0.8\linewidth]{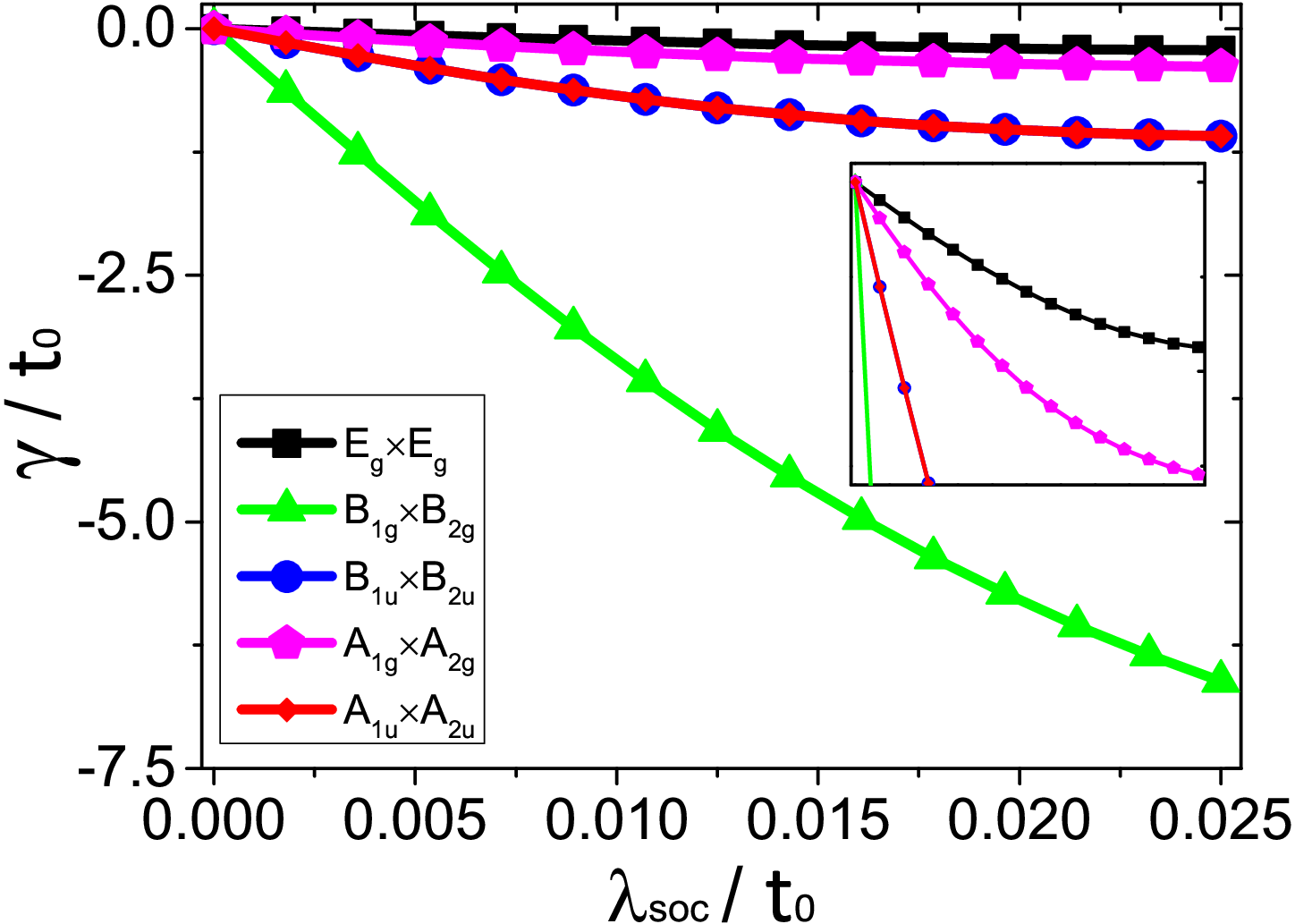}
\caption{Numerical results of $\gamma$ based on the band structure
with the same set of parameters used in Fig. \ref{fig-supp-bands}.
The Fermi energy $\mu=0.9t_0$.
$\gamma$ is nearly linearly with $\lambda_{soc}$.
}
\label{fig-sup-gamma}
\end{figure}

\section{Numerical calculations of $\gamma$}
Based on the Hamiltonian in Eq.~\eqref{supp-eq-ham0-3orb}, we numerically
calculate the Green's function, and then evaluate the Feymann diagram
discussed in the main text.
A very low temperature is used with $T=0.004t_0$, which is two orders smaller
than the Fermi energy relative to the band top.
The results are shown in Fig.~\ref{fig-sup-gamma}.
In the weak spin-orbit coupling $\lambda_{soc}$ region, $\gamma$
is linearly proportional to $\lambda_{soc}$.
The five examples for $\Delta_{1,2}$ are studied with the following
angular form factors and pairing matrices.
\begin{itemize}
	\item[1.)] Example of $A_{1g}\pm i A_{2g}$ case:
	\begin{align*}
	\Delta_1: (\cos k_x-\cos k_y) \tau_1\lambda_1 ,\; \Delta_2: (\cos k_x - \cos k_y)\tau_1\lambda_3.	
	\end{align*}
	
	\item[2.)] Example of $A_{1u}\pm i A_{2u}$ case:
	\begin{align*}
	&\Delta_1: \tau_0(\sin(k_x+k_y)\lambda_5 + \sin(k_x-k_y)\lambda_7) \\
	&\Delta_2: \tau_0(\sin(k_x+k_y)\lambda_7 - \sin(k_x-k_y)\lambda_5).		
	\end{align*}
	
	\item[3.)] Example of $B_{1g}\pm i B_{2g}$ case:
	\begin{align*}
	\Delta_1: \cos k_x\cos k_y \tau_0\lambda_1 ,\; \Delta_2: (\cos k_x + \cos k_y)\tau_1\lambda_3.	
	\end{align*}
	
	\item[4.)] Example of $B_{1u}\pm i B_{2u}$ case:
	\begin{align*}
	&\Delta_1: \tau_0(\sin(k_x+k_y)\lambda_7 + \sin(k_x-k_y)\lambda_5) \\
	&\Delta_2: \tau_0(\sin(k_x+k_y)\lambda_5 - \sin(k_x-k_y)\lambda_7).	
	\end{align*}
	
	\item[5.)] Example of $E_{g}\pm i E_{g}$ case:
	\begin{align*}
	\Delta_1: \cos (k_x + k_y) \tau_0\lambda_4 ,\; \Delta_2: \cos (k_x - k_y) \tau_0\lambda_6.	
	\end{align*}
\end{itemize}

However, we find that for the $E_u\pm i E_u$ case, the trace
operation in Eq. 4 in the main text already yields zero based
on the above three-band model, yielding $\gamma=0$, although
a non-zero $\gamma$ is allowed by symmetry.

\begin{figure}[!htbp]
\centering
\includegraphics[width=3.4in]{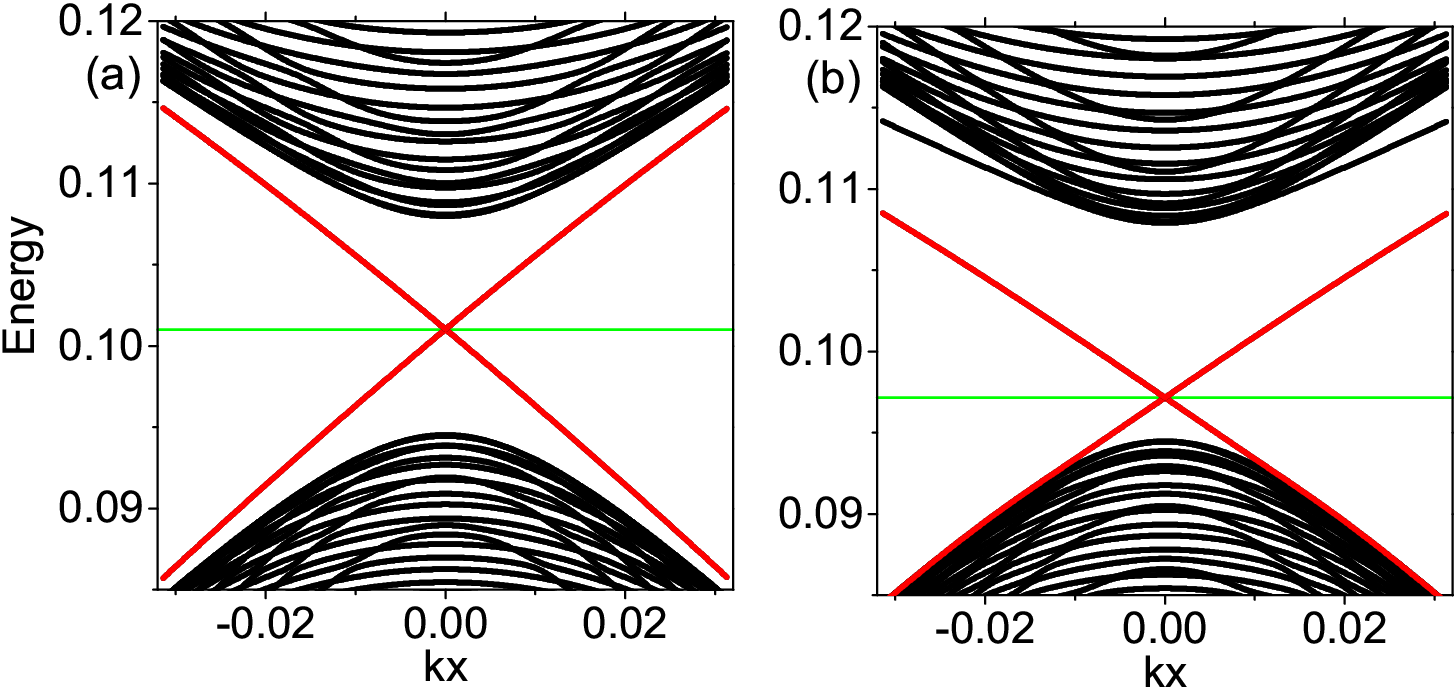}
\caption{\label{fig-neel-dirac-cone}
a) The surface Dirac cone in the absence of the Neel ordering
along the $z$-axis.
b) In the presence of the Neel ordering on both of the upper and lower surfaces, the spectra of Dirac cones
are shifted but remain non-splitting.}
\end{figure}

\section{The effect of the surface Neel order }
We discuss whether the antiferromagnetic Neel ordering at the Fe-sites
along the $z$-direction can gap out the surface Dirac cone, whose
symmetry belongs to the $B_{2u}$ representation.
It cannot be induced by mixing pairing order parameters with the same
parity in the bulk, for example, for the $s+id$-pairing, or, more precisely, $A_{1g}+ iB_{2g}$-pairing\cite{lee_prl_2009}.
Nevertheless, the inversion symmetry is broken at the surface, it could
appear at the surface for the above pairing.
Even though, the surface Dirac cone would still remain gapless due to the
protection from the mirror symmetries $\sigma_{x}$ and $\sigma_y$
along two orthogonal directions which remains in the presence of
the Neel ordering.
Hence, the ferromagnetic ordering $m_z$ is necessary to gap out
the Dirac cone.

Next we present a detailed calculation.
We define the orbital band bases of $\{\vert i,\sigma\rangle, \}$ with
$i=1\sim 4$ and $\sigma=\uparrow,\downarrow$ as follows,
\begin{align}\label{eq-basis}
\begin{split}
\vert 1 \rangle_\sigma &= \frac{1}{\sqrt{2}}\left( \phi_{x^2-y^2}^A + \phi_{x^2-y^2}^B \right) \otimes |\sigma\rangle , \\
\vert 2\rangle_\sigma &= \frac{1}{2}\left\lbrack \left(\phi_{yz}^A+\phi_{yz}^B\right) + i\left(\phi_{xz}^A+\phi_{xz}^B\right) \right\rbrack \otimes \vert \sigma\rangle , \\
\vert 3\rangle_\sigma &= \frac{1}{2}\left\lbrack \left(\phi_{yz}^A+\phi_{yz}^B\right) - i\left(\phi_{xz}^A+\phi_{xz}^B\right) \right\rbrack \otimes \vert \sigma\rangle, \\
\vert 4\rangle_\sigma &= \frac{1}{\sqrt{2}}\left( \phi_{x^2-y^2}^A - \phi_{x^2-y^2}^B \right) \otimes \vert \sigma\rangle.
\end{split}
\end{align}
where $\phi^A$ and $\phi^B$ are the three $t_{2g}$ $d$-orbitals of the
Fe atoms at $A$ and $B$-sublattices, respectively.

The Neel ordering along the $z$-direction does not have the diagonal
matrix elements in the above bases due to the vanishing of the overall
magnetization.
It does have the off-diagonal matrix elements between
$|1\rangle_\sigma$ and $|4\rangle_\sigma$, and its
Hamiltonian matrix is given by
$H_{Neel} = \begin{pmatrix} H_{Neel,\uparrow} & 0 \\ 0 & H_{Neel,\downarrow} \end{pmatrix}$, where
\begin{align}
H_{Neel,\uparrow} = N_z \begin{bmatrix}
0 & 0 & 0 & 1 \\
0 & 0 & 0 & 0 \\
0 & 0 & 0 & 0 \\
1 & 0 & 0 & 0 \\
\end{bmatrix} =  - H_{Neel,\downarrow}.
\end{align}

We find that the surface Dirac cone is still gapless in the presence of
$N_z$ by carrying out a numerical calculation for the spectra for
a 3D lattice system with the open boundary condition, shown in Fig.~\ref{fig-neel-dirac-cone}.
In fact, this two-fold degeneracy is protected by two reflection
symmetries with respect to two perpendicular mirror planes, i.e.,
$\sigma_x$ and $\sigma_y$.
When applying to half-integer spin fermions, actually $\sigma_x$ and
$\sigma_y$ anti-commute with each other, giving rise to the
protected double degeneracy.

\section{Estimation of the internal magnetic field}
In the time-reversal breaking superconducting states, there is a spontaneous magnetization in the spin channel.
Assuming each Fe site has a magnetization of $b\mu_B$, where
$\mu_B=e\hbar/2m_e c$ is the Bohr magneton, and $b$ is the relative
magnetization, then the magnetization strength $M_s$
is given by,
\begin{align}
	M_s = b \frac{\mu_B}{(r_sa_0)^3}.
\end{align}
where $a_0=\hbar^2/m_e e^2$ is the Bohr radius, $(r_sa_0)^3$ is average volume
containing one iron cation.
Therefore, the induction magnetic field $B_s$ is,
\begin{align}
\begin{split}
	B_s &= 4\pi M_s = b \frac{\alpha^2}{2r_s^3}\frac{\Phi_0}{a^2}, \\
	    &= 22.0\times\frac{b}{r_s^3}  \; \text{ Tesla}.
\end{split}	
\end{align}
where $\Phi_0=hc/2e$ is the magnetic flux quantum and $\alpha$ is the fine-structure constant.

In the FeSe$_{1-x}$Te$_{x}$ superconductors, the lattice constant
in the $a$-$b$ plane is about $a=0.379$ nm, and $c=0.596$ nm
in the $c$-direction, hence $r_s^{3}$ is estimated as $680$.
The upper bound of the relative magnetization $b$ can be estimated as
the total hole density, which is $6\pi k_F^2/(2\pi)^2 \approx 3\pi/200\approx 4.7\%$
with $k_F\approx \pi/10$ as the average Fermi momentum
of the three hole pockets (see Fig.~\ref{fig-supp-bands}).
Therefore,
\begin{align}
B_s \ll B_{s,upper}= 15 \; \text{Gauss}.
\end{align}
This estimation of the upper bound of $B_s$ is at the same order as
the saturation level of the magnetization curve of FeSe system,
which is still smaller that the lower critical field (30 Gauss)
of the superconducting FeSe.

\end{document}